\documentclass[twocolumn,floatfix]{revtex4}
	\usepackage{amssymb}
	\usepackage{amsmath}
	\usepackage{epsfig}
	\usepackage{graphicx}

	\usepackage{algorithm}
	\usepackage{algorithmic}

\begin{document}

\title{Finding local community structure in networks}
\author{Aaron Clauset}
\affiliation{Department of Computer Science, \\ University of New Mexico, Albuquerque NM 87131 \\
{\tt aaron@cs.unm.edu} }
\date{\today}

\begin{abstract}
Although the inference of global community structure in networks has recently become a topic of great interest in the physics community, all such algorithms require that the graph be completely known. Here, we define both a measure of local community structure and an algorithm that infers the hierarchy of communities that enclose a given vertex by exploring the graph one vertex at a time. This algorithm runs in time $O(k^{2}d)$ for general graphs when $d$ is the mean degree and $k$ is the number of vertices to be explored. For graphs where exploring a new vertex is time-consuming, the running time is linear, $O(k)$. We show that on computer-generated graphs this technique compares favorably to algorithms that require global knowledge. We also use this algorithm to extract meaningful local clustering information in the large recommender network of an online retailer and show the existence of mesoscopic structure.
\end{abstract}

\maketitle

\section{Introduction}

Recently, physicists have become increasingly interested in representing the patterns of interactions in complex systems as networks~\cite{Strogatz01, AB02, DM02, Newman03d}. Canonical examples include the Internet~\cite{FFF99}, the World Wide Web~\cite{Kleinberg99b}, social networks~\cite{WF94}, citation networks~\cite{Price65, Redner04} and biological networks~\cite{Ito01}. In each case, the system is modeled as a graph with $n$ vertices and $m$ edges, e.g., physical connections between computers, friendships between people and citations among academic papers.

Within these networks, the global organization of vertices into {\em communities} has garnered broad interest both inside and beyond the physics community. Conventionally, a community is taken to be a group of vertices in which there are more edges between vertices within the group than to vertices outside of it. Although the partitioning of a network into such groups is a well-studied problem, older algorithms tend to only work well in special cases~\cite{KL70, Fiedler73, PSL90, Scott00, Newman04b}. Several algorithms have recently been proposed within the physics community, and have been shown to reliably extract known community structure in real world networks~\cite{GN02, NG04, newman-fast, Radicchi04, WH04a, clauset-newman-fast}. Similarly, the computer science community has proposed algorithms based on the concept of flow~\cite{Flake02}.

However, each of these algorithms require knowledge of the entire structure of the graph. This constraint is problematic for networks like the World Wide Web, which for all practical purposes is too large and too dynamic to ever be known fully, or networks which are larger than can be accommodated by the fastest algorithms~\cite{clauset-newman-fast}. In spite of these limitations, we would still like to make quantitative statements about community structure, albeit confined to some accessible and known region of the graph in question. For instance, we might like to quantify the local communities of either a person given their social network, or a particular website given its local topology in the World Wide Web.

Here, we propose a general measure of local community structure, which we call {\em local modularity}, for graphs in which we lack global knowledge and which must be explored one vertex at a time. We then define a fast agglomerative algorithm that maximizes the local modularity in a greedy fashion, and test the algorithm's performance on a series of computer-generated networks with known community structure. Finally, we use this algorithm to analyze the local community structure of the online retailer Amazon.com's recommender network, which is composed of more than $400~000$ vertices and $2$ million edges. Through this analysis, we demonstrate the existence of mesoscopic network structure that is distinct from both the microstructure of vertex statistics and the global community structure previously given in~\cite{clauset-newman-fast}. Interestingly, we find a wide variety of local community structures, and that generally, the local modularity of the network surrounding a vertex is negatively correlated with its degree.

\section{Local Modularity}
The inference of community structure can generally be reduced to identifying a partitioning of the graph that maximizes some quantitative notion of community structure. However, when we lack global knowledge of the graph's topology, a measure of community structure must necessarily be independent of those global properties. For instance, this requirement precludes the use of the modularity metric $Q$, due to Newman and Girvan~\cite{NG04}, as it depends on $m$.

Suppose that in the graph~$\mathcal{G}$, we have perfect knowledge of the connectivity of some set of vertices, i.e., the known portion of the graph, which we denote~$\mathcal{C}$. This necessarily implies the existence of a set of vertices $\mathcal{U}$ about which we know only their adjacencies to $\mathcal{C}$. Further, let us assume that the only way we may gain additional knowledge about $\mathcal{G}$ is by visiting some neighboring vertex $v_{i}\in\mathcal{U}$, which yields a list of its adjacencies. As a result, $v_{i}$ becomes a member of~$\mathcal{C}$, and additional unknown vertices may be added to~$\mathcal{U}$. This vertex-at-a-time discovery process is directly analogous to the manner in which ``spider'' or ``crawler'' programs harvest the hyperlink structure of the World Wide Web.

The adjacency matrix of such a partially explored graph is given by
\begin{equation}
A_{ij} = \left\{\begin{array}{ll}
           1 & \quad\mbox{if vertices $i$ and $j$ are connected,} \\
               & \quad\mbox{and either vertex is in $\mathcal{C}$}\\
           0 & \quad\mbox{otherwise.}
         \end{array}
         \right.
\end{equation}
If we consider~$\mathcal{C}$ to constitute a local community, the most simple measure of the quality of such a partitioning of~$\mathcal{G}$ is simply the fraction of known adjacencies that are completely internal to~$\mathcal{C}$. This quantity is given by
\begin{equation}
\frac{\sum_{ij} A_{ij}\xi(i,j)}{\sum_{ij} A_{ij} } = \frac{1}{2m^{*}}\sum_{ij}A_{ij}\xi(i,j) \enspace ,
\end{equation}
where $m^{*}=\frac{1}{2}\sum_{ij}A_{ij}$, the number of edges in the partial adjacency matrix, and $\xi(i,j)$ is $1$ if both $v_{i}$ and $v_{j}$ are in~$\mathcal{C}$ and $0$ otherwise. This quantity will be large when~$\mathcal{C}$ has many internal connections, and few connections to the unknown portion of the graph. This measure also has the property that when $|\mathcal{C}|\gg|\mathcal{U}|$, the partition will almost always appear to be good.

\begin{figure} [t]
\begin{center}
\includegraphics[scale=0.75]{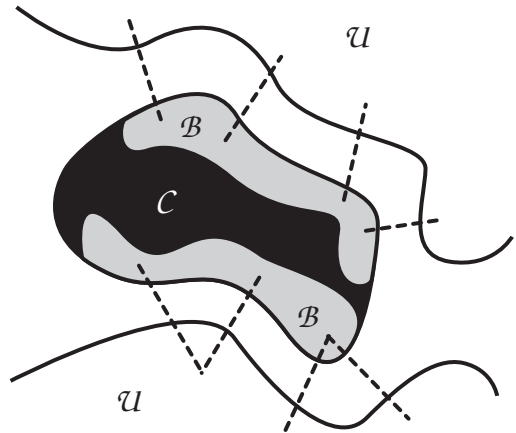}
\caption{An illustration of the division of an abstract graph into the local community $\mathcal{C}$, its boundary $\mathcal{B}$ and the edges which connect $\mathcal{B}$ to the largely unknown neighbors $\mathcal{U}$. }
\label{fig:diagram}
\end{center}
\end{figure}

If we restrict our consideration to those vertices in the subset of $\mathcal{C}$ that have at least one neighbor in $\mathcal{U}$, i.e., the vertices which make up the {\em boundary} of $\mathcal{C}$ (Fig.~\ref{fig:diagram}), we obtain a direct measure of the sharpness of that boundary. Additionally, this measure is independent of the size of the enclosed community. Intuitively, we expect that a community with a sharp boundary will have few connections from its boundary to the unknown portion of the graph, while having a greater proportion of connections from the boundary back into the local community  In the interest of keeping the notation concise, let us denote those vertices that comprise the boundary as $\mathcal{B}$, and the boundary-adjacency matrix as
	\begin{equation}
	B_{ij} = \left\{\begin{array}{ll}
	           1 & \quad\mbox{if vertices $i$ and $j$ are connected,} \\
	               & \quad\mbox{and either vertex is in $\mathcal{B}$}\\
	           0 & \quad\mbox{otherwise.}
	         \end{array}\right.
	\end{equation}
Thus, we define the local modularity $R$ to be
	\begin{equation}
	R = \frac{\sum_{ij} B_{ij}\delta(i,j)}{\sum_{ij} B_{ij} } = \frac{I}{T} \enspace ,
	\label{eq:local}
	\end{equation}
where $\delta(i,j)$ is $1$ when either $v_{i}\in\mathcal{B}$ and $v_{j}\in\mathcal{C}$ or vice versa, and is $0$ otherwise. Here, $T$ is the number of edges with one or more endpoints in $\mathcal{B}$, while $I$ is the number of those edges with neither endpoint in $\mathcal{U}$. This measure assumes an unweighted graph, although the weighted generalization is straightforward~\cite{Newman04}.

A few comments regarding this formulation are worthwhile before proceeding. By considering the fraction of boundary edges which are internal to $\mathcal{C}$, we ensure that our measure of local modularity lies on the interval $0< R<1$, where its value is directly proportional to sharpness of the boundary given by $\mathcal{B}$. This is true except when the entire component has been discovered, at which point $R$ is undefined. If we like, we may set $R=1$ in that case in order to match the intuitive notion that an entire component constitutes the strongest kind of community. Finally, there are certainly alternative measures that can be defined on $\mathcal{B}$, however, in this paper we consider only the one given. 

\section{The Algorithm}
For graphs like the World Wide Web, in which one must literally crawl the network in order to discover the adjacency matrix, any analysis of local community structure must necessarily begin at some source vertex $v_{0}$. In general, if the explored portion of the graph has $k$ vertices, the number of ways to partition it into two sets, those vertices considered a part of the same local community as the source vertex and those considered outside of it, is given by $2^{k-2}-1$, which is exponential in the size of the explored portion of the network. In this section, we describe an algorithm that only takes time polynomial in $k$, and that infers local community structure by using the vertex-at-a-time discovery process subject to maximizing our measure of local modularity.

Initially, we place the source vertex in the community, $v_{0} = \mathcal{C}$, and place its neighbors in~$\mathcal{U}$. At each step, the algorithm adds to $\mathcal{C}$ (and to $\mathcal{B}$, if necessary) the neighboring vertex that results in the largest increase (or smallest decrease) in $R$, breaking ties randomly.
Finally, we add to $\mathcal{U}$ any newly discovered vertices, and update our estimate of $R$. This process continues until it has agglomerated either a given number of vertices $k$, or it has discovered the entire enclosing component, whichever happens first. Pseudocode for this process is given in Algorithm 1. As we will see in the two subsequent sections, this algorithm performs well on both computer-generated graphs with some known community structure and on real world graphs. 

\begin{algorithm}[t]
\label{alg:local}
\caption{The general algorithm for the greedy maximization of local modularity, as given in the text.}
\begin{algorithmic}
\STATE add $v_{0}$ to $\mathcal{C}$
\STATE add all neighbors of $v_{0}$ to $\mathcal{U}$
\STATE set $\mathcal{B}=v_{0}$
\WHILE{$|\mathcal{C}| < k$} 
\FOR{each $v_{j}\in \mathcal{U}$}
\STATE compute $\Delta R_{j}$
\ENDFOR
\STATE find $v_{j}$ such that $\Delta R_{j}$ is maximum
\STATE add that $v_{j}$ to $\mathcal{C}$
\STATE add all new neighbors of that $v_{j}$ to $\mathcal{U}$
\STATE update $R$ and $\mathcal{B}$
\ENDWHILE
\end{algorithmic}
\end{algorithm}


The computation of the $\Delta R_{j}$ associated with each $v_{j}\in\mathcal{U}$ can be done quickly using an expression derived from equation (\ref{eq:local}):
\begin{equation}
\Delta R_{j} = \frac{x - Ry-z(1-R)}{T - z + y}\enspace ,
\end{equation}
where $x$ is the number of edges in $T$ that terminated at $v_{j}$, $y$ is the number of edges that will be added to $T$ by the agglomeration of $v_{j}$ (i.e., the degree of $v_{j}$ is $x+y$), and $z$ is the number of edges that will be removed from $T$ by the agglomeration. Because $\Delta R_{j}$ depends on the current value of $R$, and on the $y$ and $z$ that correspond to $v_{j}$, each step of the algorithm takes time proportional to the number of vertices in $\mathcal{U}$. This is roughly $kd$, where $d$ is the mean degree of the graph; we note that this will be a significant overestimate for graphs with non-trivial clustering coefficients, significant community structure, or when $\mathcal{C}$ is a large portion of the graph. Thus, in general, the running time for the algorithm is $O(k^{2}d)$, or simply $O(k^{2})$ for a sparse graph, i.e., when $m\sim n$. As it agglomerates vertices, the algorithm outputs a function $R(t)$, the local modularity of the community centered on $v_{0}$ after $t$ steps, and a list of vertices paired with the time of their agglomeration.

The above calculation of the running time is somewhat misleading as it assumes that the algorithm is dominated by the time required to calculate the $\Delta R_{j}$ for each vertex in $\mathcal{U}$; however, for graphs like the World Wide Web, where adding a new vertex to $\mathcal{U}$ requires the algorithm to fetch a web page from a remote server, the running time will instead be dominated by the time-consuming retrieval. When this is true, the running time is linear in the size of the explored subgraph, $O(k)$.

A few comments regarding this algorithm are due. Because of the greedy maximization of local modularity, a neighboring high degree vertex will not be agglomerated until the number of its unknown neighbors has decreased sufficiently. It is this behavior that allows the algorithm to avoid crossing a community boundary until absolutely necessary. Additionally, the algorithm is somewhat sensitive to the degree distribution of the source vertex's neighbors: when the source degree is high, the algorithm will first explore its low degree neighbors. This implicitly assumes that high degree vertices are likely to sit at the boundary of several local communities. While certainly not the case in general, this may be true for some real world networks. We shall return to this idea in a later section.

Finally, although one could certainly stop the algorithm once the first enclosing community has been found, in principle, there is no reason that it cannot continue until some arbitrary number of vertices have been agglomerated. Doing so yields the hierarchy of communities which enclose the source vertex. In a sense, this process is akin to the following: given the dendrogram of the global community hierarchy, walk upward toward the root from some leaf $v_{0}$ and observe the successive hierarchical relationships as represented by junctions in the dendrogram. In that sense, the enclosing communities inferred by our algorithm for some source vertex is the community hierarchy {\em from the perspective of that vertex}. 

\section{computer-generated Graphs}
As has become standard with testing community inference techniques, we apply our algorithm to a set of computer-generated random graphs which have known community structure~\cite{NG04}. In these graphs, $n=128$ vertices are divided into four equal-sized communities of $32$ vertices. Each vertex has a total expected degree $z$ which is divided between intra- and inter-community edges such that  $z=z_{in}+z_{out}$. These edges are placed independently and at random so that, in expectation, the values of $z_{in}$ and $z_{out}$ are respected. By holding the expected degree constant $z=16$, we may tune the sharpness of the community boundaries by varying $z_{out}$. Note that for these graphs, when $z_{out}=12$, edges between vertices in the same group are just as likely as edges between vertices that are not.

\begin{figure} [t]
\begin{center}
\includegraphics[scale=0.45]{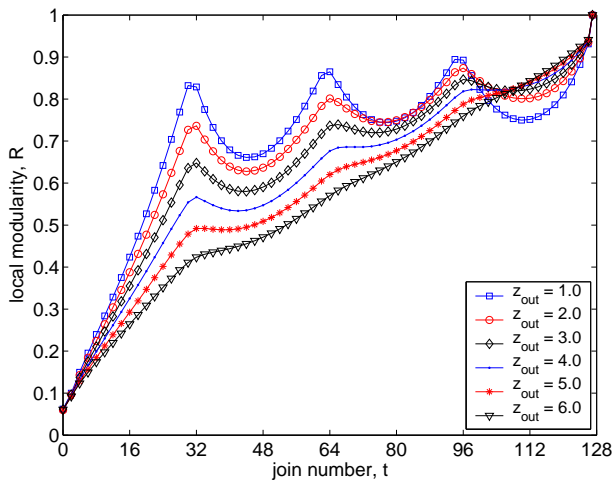}
\caption{Local modularity $R$ as a function of the number of steps $t$, averaged over $500$ computer-generated networks as described in the text; error bars are omitted for clarity. By varying the expected number of inter-community edges per node $z_{out}$, the strength of the community boundaries are varied. }
\label{fig:modularity}
\end{center}
\end{figure}

Figure~\ref{fig:modularity} shows the average local modularity $R$ as a function of the number of steps $t$, over $500$ realizations of the graphs described above. For the sake of clarity, only data series for $z_{out}\leq6.0$ are shown and error bars are omitted. Sharp community boundaries correspond to peaks in the curve. As $z_{out}$ grows, the sharpness of the boundaries and the height of the peaks decrease proportionally. When the first derivative is positive everywhere, e.g., for $z_{out}> 5$, the inferred locations of the community boundaries may be extracted by finding local minima in the second derivative, possibly after some smoothing.

\begin{figure} [t]
\begin{center}
\includegraphics[scale=0.45]{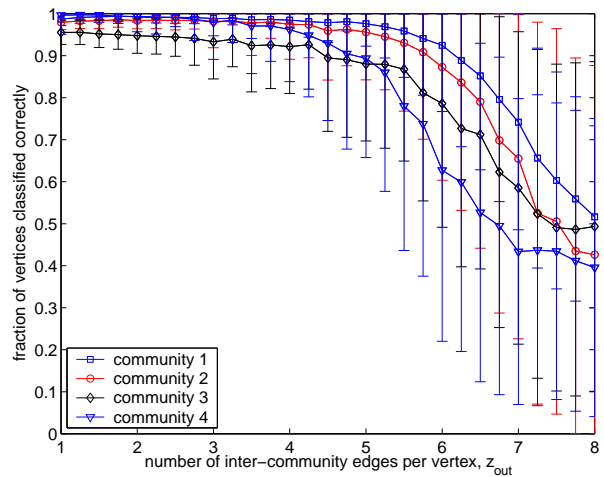}
\caption{Fraction of correctly classified nodes, by community, as a function of the number of inter-community edges $z_{out}$. Although there the variance increases as the community boundaries become less sharp, the average behavior (over $500$ realizations) degrades gracefully, and compares favorably with methods which use global information. }
\label{fig:fraction}
\end{center}
\end{figure}

From this information we may grade the performance of the algorithm on the computer-generated graphs.  Figure~\ref{fig:fraction} shows the average fraction of correctly classified vertices for each of the four communities as a function of $z_{out}$, over $500$ realizations; error bars depict one standard deviation. As a method for inferring the first enclosing community, our algorithm classifies more than $50\%$ of the vertices correctly even when the boundaries are weak, i.e., when $z_{out}=8$. Although the variance in the quality of classification grows as $z_{out}$ approaches $z_{in}$, this is to be expected given that the algorithm uses only local information for its inference, and large local fluctuations may mislead the algorithm. For computer-generated graphs such as these, the performance of our algorithm compares favorably to that of more global methods~\cite{NG04,newman-fast,Radicchi04}.

Recently, another approach to inferring community structure using only local information appeared~\cite{Bagrow04}. This alternative technique relies upon growing a breadth-first tree outward from the source vertex $v_{0}$, until the rate of expansion falls below an arbitrary threshold. The uniform exploration has the property that some level in the tree will correspond to a good partitioning only when $v_{0}$ is equidistant from all parts of its enclosing community's boundary. On the other hand, by exploring the surrounding graph one vertex at a time, our algorithm will avoid crossing boundaries until it has explored the remainder of the enclosing community.

\section{Local Co-Purchasing Habits}
In this section, we apply our local inference algorithm to the recommender network of Amazon.com, collected in \mbox{August 2003}, which has $n=409~687$ vertices, $m=2~464~630$ edges and thus a mean degree of $12.03$. We note that the degree distribution is fairly right-skewed, having a standard deviation of $14.64$. Here, vertices are items such as books and digital media sold on Amazon's website, while edges connect pairs of items that are frequently purchased together by customers. It is this co-purchasing data that yields recommendations for customers as they browse the online store. Although in general, the algorithm we have described is intended for graphs like the World Wide Web, the Amazon recommender network has the advantage that, by virtue of being both very large and fully known, we may explore global regularities in local community structure without concern for sampling bias in the choice of source vertices. Additionally, we may check the inferred the community structures against our, admittedly heuristic, notions of correctness. 

\begin{figure} [t]
\begin{center}
\includegraphics[scale=0.45]{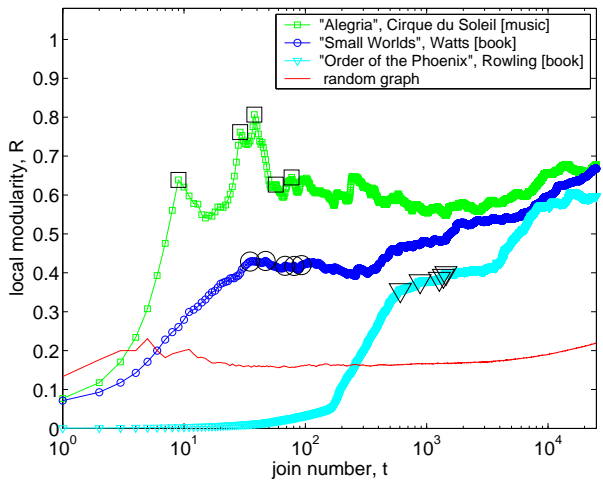}
\caption{Local modularity $R$ for three items in the Amazon.com recommender network, shown on log-linear axes. For comparison, the time series for a random graph with the same degree distribution is shown. The large open symbols indicate the locations of the five strongest enclosing communities. }
\label{fig:amazon}
\end{center}
\end{figure}

As illustrative examples, we choose three qualitatively different items as source vertices: the compact disc {\em Alegria} by \mbox{Cirque du Soleil}, the book \mbox{{\em Small Worlds}} \mbox{by Duncan Watts}, and the book {\em Harry Potter and the Order of the Phoenix} by \mbox{J.K. Rowling}. These items have degree $15$, $19$ and $3117$ respectively. At the time the network data was collected, the Harry Potter book was the highest degree vertex in the network, its release date having been June 2003. For each of these items, we explore $k=25~000$ associated vertices. Figure~\ref{fig:amazon} illustrates the local modularity as a function of the number of steps $t$ for each item; an analogous data series for a random graph with the same degree distribution~\cite{Bollobas95} has been plotted for comparison. We mark the locations of the five principle enclosing communities with large open symbols.

These time series have several distinguishing features. First, {\em Alegria} has the smallest enclosing communities, composed of $t=\{10, 30, 39, 58, 78\}$ vertices, and these communities are associated with high values of local modularity. The first five enclosing communities all have $R>0.62$, while the third community corresponds to $R=0.81$, indicating that only about $20\%$ of boundary edges reach out to the rest of the network. In contrast, the communities of \mbox{{\em Small Worlds}} contain $t=\{36, 48, 69, 82, 94\}$ vertices, while the Harry Potter book's communities are extremely large, containing $t=\{607, 883, 1270, 1374, 1438\}$ vertices. Both sets have only moderate values of local modularity, $R\leq0.43$. It is notable that the local modularity functions for all three items follow relatively distinct trajectories until the algorithm has agglomerated roughly $10~000$ items. Beyond that point, the curves begin to converge, indicating that, from the perspectives of the source vertices, the local community structure has become relatively similar. 

\begin{figure} [t]
\begin{center}
\includegraphics[scale=0.43]{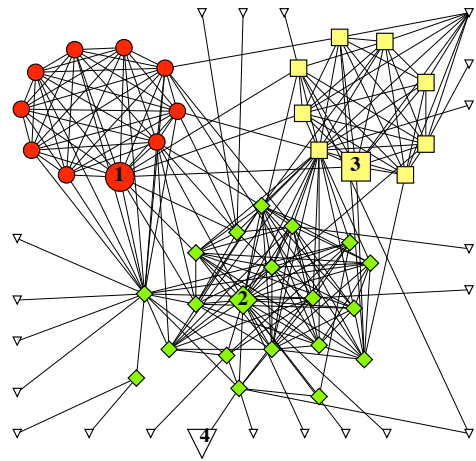}
\caption{The first three enclosing communities for Cirque du Soleil's {\em Alegria} in Amazon.com's recommender network; communities are distinguished by shape (circles, diamonds, squares respectively). Connections to triangles represent connections to items in the remaining unknown portion of the graph. {\em Alegria} and {\em Order of the Phoenix} are denoted by $1$ and $4$ respectively. }
\label{fig:subgraph}
\end{center}
\end{figure}

To illustrate the inferred local structure, we show the partial subgraph that corresponds to the first three enclosing local communities for the compact disc {\em Alegria} in Figure~\ref{fig:subgraph}. Here, communities are distinguished by shape according to the order of discovery (circle, diamond and square respectively), and vertices beyond these communities are denoted by triangles. Items in the first enclosing community are uniformly compact discs produced by Cirque du Soleil. Items in the second are slightly more diverse, including movies and books about the troupe, the Cirque du Soleil compact disc entitled {\em Varekai}, and one compact disc by a band called Era; the third group contains both new and old Cirque du Soleil movies. {\em Varekai} appears to have been placed outside the first community because it has fewer connections to those items than to items in the subsequent enclosing communities. Briefly, we find that the enclosing local communities of \mbox{{\em Small Worlds}} are populated by texts in sociology and social network analysis, while the Harry Potter book's communities have little topical similarity.

In Figure~\ref{fig:subgraph}, the labels $1$ and $4$ denote the items {\em Alegria} and {\em Order of the Phoenix}, respectively. It is notable that these items are only three steps away in the graph, yet have extremely different local community structures (Fig.~\ref{fig:amazon}). If an item's popularity is reflected by its degree, then it is reasonable to believe that the strength of the source vertex's local community structure may be inversely related to its degree. That is, popular items like {\em Order of the Phoenix} may tend to link many well-defined communities by virtue of being purchased by a large number of customers with diverse interests, while niche items like Cirque du Soleil's {\em Alegria} exhibit stronger local community structure as the result of more specific co-purchasing habits. Such structure appears to be distinct from both the macroscopic structure discovered using global community inference methods~\cite{clauset-newman-fast}, and the microscopic structure of simple vertex-statistics such as clustering or assortativity.

\begin{figure} [t]
\begin{center}
\includegraphics[scale=0.45]{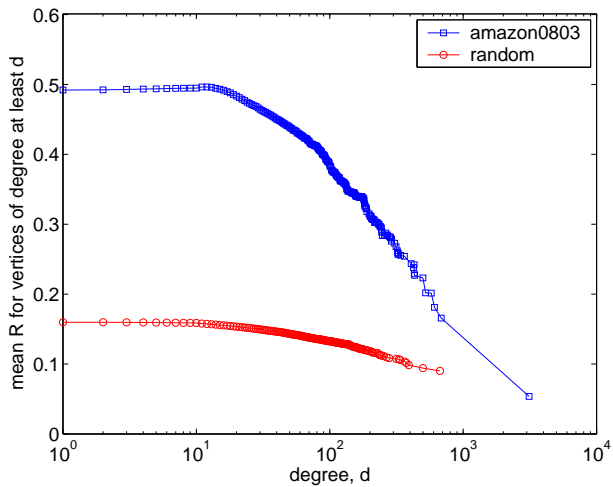}
\caption{The average local modularity over the first $250$ steps for source vertices with degree at least $d$. The ``knee'' in the upper data series is located at $d=13$; the mean degree for the network is $12.03$. The logarithmic falloff illustrates the negative correlation between source vertex degree and the strength of the surrounding local community. }
\label{fig:cumulative}
\end{center}
\end{figure}

With the exception of social networks, the degree of adjacent vertices appears to be negatively correlated in most networks. This property is often called ``disassortative'' mixing~\cite{Newman03b}, and can be caused by a high clustering coefficient, global community structure or a specific social mechanism~\cite{Newman03}. However, for the Amazon recommender network, we find that the assortativity coefficient is not statistically different from zero, $r = -3.01\times 10^{-19}\pm 1.49\times 10^{-4}$, yet the network exhibits a non-trivial clustering coefficient, $c = 0.17$ and strong global community structure structure with a peak modularity of $Q=0.745$~\cite{clauset-newman-fast}. Returning to the suggestion above that there is an inverse relationship between the degree of the source vertex and the strength of its surrounding community structure, we sample for $100~000$ random vertices the average local modularity over the first $k=250$ steps. We find the average local modularity to be relatively high, $\bar{R}_{amzn}=0.49\pm0.08$, while a random graph with the same degree distribution yields $\bar{R}_{rand} = 0.16\pm0.01$. The variance for the Amazon graph is due to the contributions of high degree vertices. In Figure~\ref{fig:cumulative}, we plot from our random sample, the average local modularity for all source vertices with degree at least $d$. Notably, the average is relatively constant until $d=13$, after which it falls off logarithmically. This supports the hypothesis that, in the recommender network, there is a weak inverse relationship between the degree of the source vertex and the strength of its surrounding local community.

Naturally, there are many ways to use the concept of local community structure to understand the mesoscopic properties of real world networks. Further characterizations of the Amazon graph are beyond the scope of this paper, but we propose a rigorous exploration of the relationship between the source vertex degree and its surrounding local community structure as a topic for future work.



\section{Conclusions}
Although many recent algorithms have appeared in the physics literature for the inference of community structure when the entire graph structure is known, there has been little consideration of graphs that are either too large for even the fastest known techniques, or that are, like the World Wide Web, too large or too dynamic to ever be fully known. Here, we define a measure of community structure which depends only on the topology of some known portion of a graph. We then give a simple fast, agglomerative algorithm that greedily maximizes our measure as it explores the graph one vertex at a time. When the time it takes to retrieve the adjacencies of a vertex is small, this algorithm runs in time $O(k^{2}d)$ for general graphs when it explores $k$ vertices and the graph has mean degree $d$. For sparse graphs, i.e., when $m\sim n$, this is simply $O(k^{2})$. On the other hand, when visiting a new vertex to retrieve its adjacencies dominates the running time, e.g., downloading a web page on the World Wide Web, the algorithm takes time linear in the size of the explored subgraph, $O(k)$. Generally, if we are interested in making quantitative statements about local structure, that is, when $k\ll n$, it is much more reasonable to use an algorithm which is linear or even quadratic in $k$, than an algorithm that is linear in the size of the graph $n$. Finally, we note that our algorithm's simplicity will make it especially easy to incoporate into web spider or crawler programs for the discovery of local community structures on the World Wide Web graph.

Using computer-generated graphs with known community structure, we show that our algorithm extracts this structure and that its performance compares favorably with other community structure algorithms~\cite{NG04,newman-fast,Radicchi04} that rely on global information. We then apply our algorithm to the large recommender network of the online retailer Amazon.com, and extract the local hierarchy of communities for several qualitatively distinct items. We further show that a vertex's degree is inversely related to the strength of its surrounding local structure. This discovery points to the existence of mesoscopic topological regularities that have not been characterized previously. Finally, this algorithm should allow researchers to characterize the structure of a wide variety of other graphs, and we look forward to seeing such applications.

\acknowledgements{The author is grateful to Cristopher Moore for his persistent encouragement and guidance, Mark Newman for helpful discussions about community structure, and to Amazon.com and Eric Promislow for providing the recommender network data. This work was supported in part  by the National Science Foundation under grants PHY-0200909 and ITR-0324845.}


\end{document}